\documentclass[aps,floats,prl,showpacs,twocolumn ]{revtex4}

\newcommand{\ha}{{\hat \alpha}}

\usepackage{amsfonts,amsmath,mathrsfs} \usepackage{bm} \usepackage{dcolumn}
\usepackage{epsfig} \usepackage{latexsym}

\begin{document}
\title{Confinement-deconfinement interplay in quantum phases of doped Mott insulators}

\author{Peng Ye$^{1}$, Chu-Shun Tian$^{1,2}$, Xiao-Liang Qi$^{3,4}$ and Zheng-Yu Weng$^1$}

\affiliation{$^{1}$ Institute for Advanced Study, Tsinghua
University,
 Beijing, 100084, P. R. China\\
$^{2}$ Institut f\"{u}r Theoretische Physik, Universit\"{a}t zu
 K\"{o}ln, D-50937 K\"{o}ln, Germany \\
$^{3}$ Microsoft Research, Station Q, Elings Hall, University of
California, Santa Barbara, CA 93106, USA\\
$^{4}$ Department of Physics, Stanford University, Stanford, CA
94305, USA}
\begin{abstract}
{\rm It is generally accepted that doped Mott insulators can be well
characterized by the $t$-$J$ model. In the $t$-$J$ model, the
electron fractionalization is dictated by the phase string effect.
We found that in the underdoped regime, the antiferromagnetic and
superconducting phases are dual: in the former, holons are confined
while spinons are deconfined, and {\it vice versa} in the latter.
These two phases are separated by a novel phase, the so-called
Bose-insulating phase, where both holons and spinons are deconfined.
A pair of Wilson loops was found to constitute a complete set of
order parameters determining this zero-temperature phase diagram.
The quantum phase transitions between these phases are suggested to
be of non-Landau-Ginzburg-Wilson type.}
\end{abstract}


\maketitle

{\it Introduction.}--The concept of fractionalization, finding its
analog
in quantum chromodynamics, is nowadays a guiding principle of
strongly correlated systems \cite{Anderson87}. Specifically, a
quasiparticle (like electron), at short spacetime scales, is
effectively fractionalized into a few degrees of freedom (like spin
and charge) which, at
large
scales, are ``glued'' by emergent gauge degrees of freedom. As such,
the gauge fields affect profoundly the formation of various
novel quantum phases \cite{Lee06}. A fundamental issue is the nature
of the quantum phase transition between these unconventional phases.
In the conventional Landau-Ginzburg-Wilson (LGW) symmetry-breaking
paradigm, different phases are characterized by the presence or
absence of certain local order parameters. On the contrary, the
fluctuation of emergent gauge field may contribute to low energy
excitations of the system. In general, the transition between such
phases  cannot be described by the symmetry breaking paradigm, as is
exemplified in Ref.~\cite{Fisher04}.

Practically, an important playground of fractionalization is doped
cuprates \cite{Anderson87,Lee06}. Despite the debate about the
nature of the cuprate phase diagram, it is widely accepted that the
essential physics of high $T_c$ superconductivity is captured by a
doped Mott insulator where the electron fractionalization into
spinons and holons is seemingly inevitable
\cite{Anderson87,Lee06,Fisher00,Sachdev03}. Indeed, there have been
abundant evidence indicating that the electron fractionalization
leads to new quantum phases \cite{Lee06} and their transitions may
not be understood in terms of symmetry breaking. This opinion has
been reiterated very recently in Ref.~\cite{Zaanen09}, where it was
conjectured that the drastic change in the nature of quantum
statistics--a direct manifestation of electron fractionalization--is
at the root of the pseudogap phase found in the cuprates.

Many fundamental issues arise thereby. {\it How does the electron
fractionalization turn an antiferromagnet into a superconductor upon
doping? What does the phase diagram look like and what are the
underlying order parameters?} Substantial efforts
\cite{Weng97,Weng07,KQW05,Weng02,Weng10} suggest that an exact
non-perturbative result, the so-called {\it phase string effect},
discovered for the $t$-$J$ model \cite{Weng97} is at the core of
these issues. In this Letter, we present an analytical study of
these open issues based on this exact result. We stress that our
theory may be extended to other systems such as the Hubbard model on
the honeycomb lattice currently undergoing intense investigations
\cite{Xu09}.


\begin{figure}[h]
  \centering
 \includegraphics[width=8cm]{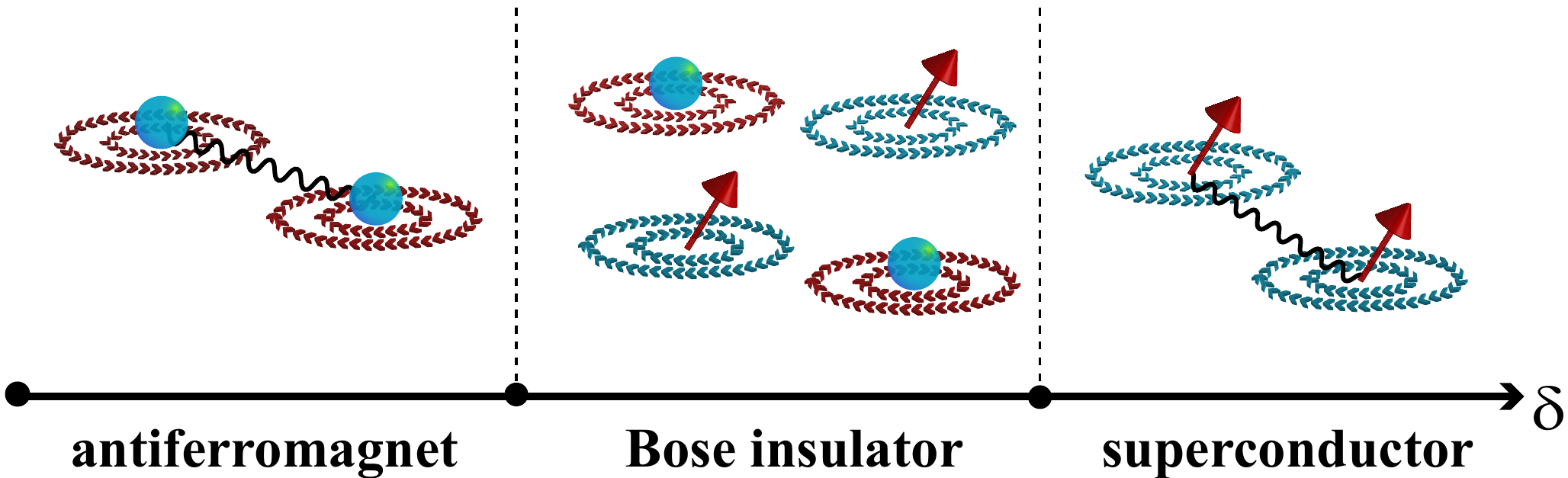}
  \caption{\label{mainresult} (Color online) Zero temperature phase diagram of underdoped Mott insulators.
  The ball (in blue) and the arrow (in red) stand for the holon and spinon, respectively.
  The vortex, in red (blue), surrounding
  a holon (spinon) arises from the spinon (holon) condensate. The wavy line stands for the confinement.}
\end{figure}

{\it Main results and qualitative discussions.}--In essence, the
phase string effect renders an electron ``fractionalized'' into two
topological objects, the spinon and the holon, each of which is
bosonic and carries a $\pi$-flux \cite{unit}. A $U(1)$ gauge field
$A^{h}_\mu$ ($A^{s}_\mu$), radiated by the holons (spinons),
interacts with spinons (holons) through minimal coupling. (This is
the so-called mutual statistical interaction which was also found in
different contexts \cite{Xu09,Galitskii05}). The macroscopic
electric current (density), ${\bf j}^h$, is fully carried by holons
and driven by both the external electric field ${\bf E}$ and the
``electric field'' ${\bf E}^s$ resulting from $A^s_\mu$ induced by
spinons. Interestingly, ${\bf E}^s$ finds its origin analogous to
that of Ohmic dissipation in type-II superconductors: each spinon
mimics a ``magnetic vortex'' suspending in holon fluids and, upon
moving, generates an electric field {\it antiparallel} to ${\bf
j}^h$, i.e., ${\bf E}^s=-\pi^2 \sigma_s {\bf j}^h$, with $\sigma_s$
the spinon conductivity characterizing the mobility of spinons. From
the Ohm's law, i.e., $\sigma_h^{-1}{\bf j}^h=-\pi^2 \sigma_s {\bf
j}^h+{\bf E}$, with $\sigma_{h}$ the holon conductivity, we find
that the (electric) resistivity, defined as $\sigma^{-1} {\bf j}^h
={\bf E}$, is
\begin{eqnarray}
\sigma^{-1}=\sigma_h^{-1} + \pi^2 \sigma_s. \label{conductivity}
\end{eqnarray}
A microscopic derivation of this composition rule will be given
later.

For the doping $\delta$ sufficiently small, the antiferromagnetic
(AF) phase is reached where the spinons are deconfined and form
superfluids, i.e., $\sigma_s\rightarrow \infty$. According to
Eq.~(\ref{conductivity}) this phase is insulating, giving a
vanishing $\sigma$. Indeed, single holon cannot appear in the
excitation spectrum. Rather, holons are excited in pair and are
logarithmically confined. These pairs are bound to the vortices of
spinon superfluids and are thus immobile. Upon increasing $\delta$,
the typical size of holon pairs becomes larger and larger and, as a
result, the confinement becomes weaker and weaker. Eventually, a
quantum critical point (QCP) is reached. Beyond this QCP holons
become deconfined. For sufficiently large $\delta$, a dual scenario
applies. The holons are deconfined and form superfluids, i.e.,
$\sigma_h\rightarrow \infty$. In this phase, single spinon cannot be
excited. Rather, spinons are excited in pair, logarithmically
confined and bound to the vortices of holon superfluids, i.e.,
$\sigma_s=0$. (Consequently, the excitation spectrum is composed of
integer spin excitations.) According to Eq.~(\ref{conductivity})
this phase is superconducting (SC), i.e., $\sigma\rightarrow
\infty$. Upon decreasing $\delta$, the confinement becomes weaker
and weaker. Eventually, another QCP is reached: for smaller $\delta$
spinons are deconfined. Thus, an intermediate phase can appear where
both spinon and holon vortices are condensed. In this phase,
$\sigma_{s,h}\rightarrow \infty$ implies $\sigma=0$. This brings us
to the term, Bose insulating (BI) phase.

The zero temperature phase diagram described above is summarized in
Fig.~\ref{mainresult}. All three phases are characterized by
confinement or deconfinement of spinons and holons, formally
implemented by unconventional order parameters--a pair of Wilson
loops $W_\delta^{s,h}[{\cal C}]$ with ${\cal C}$ a spacetime
rectangle. Specifically, they are defined as the expectation values
of $e^{i\oint_{{\cal C}} A^{s,h}_\mu dx^\mu}$ that probes the
interaction between a pair of test holons (spinons).
$W_\delta^{s}[{\cal C}]$ ($W_\delta^{h}[{\cal C}]$) displays
nonanalyticity at the holon (spinon) deconfinement QCP.
The existence of these nonlocal ``order parameters'' suggests the
non-LGW nature of the quantum phase transitions: in contrast to an
LGW type scenario of order parameter competing, the AF and SC phases
are intrinsically incompatible such that they cannot be turned into
unless confinement or deconfinement occurs. In particular, they are
generally separated by the BI phase, with the AF (SC)-BI transition
as a holon (spinon) deconfinement QCP. We now turn to present some
technical details.


{\it Lattice field theory.}--Formally, we start from the Hamiltonian
of the $t$-$J$ model, which consists of the superexchange and
hopping terms, describing the spin-flip and charge hopping process,
respectively. An exact transformation \cite{Weng97} keeping track of
the phase string effect transforms the superexchange term into
($\alpha=1,2$)
\begin{eqnarray}
H_J \propto
\sum_{i\alpha\sigma\sigma'}e^{-i\sigma
A_\alpha^h(i)}b_{i-\sigma}^\dagger b_{i+\ha \sigma}^\dagger
e^{i\sigma' A_\alpha^h(i)}b_{i+\ha \sigma'} b_{i-\sigma'} \nonumber
\end{eqnarray}
and the hopping term into
\begin{eqnarray} H_t \propto
\sum_{i \alpha \sigma} e^{i A_\alpha^s(i)} h_i^\dagger h_{i+\ha}
e^{-i\sigma A_\alpha^h(i)} b_{i+\ha \sigma}^\dagger b_{i\sigma} +
{\rm h. c.}, \nonumber
\end{eqnarray}
with the coefficients omitted and ${\rm h.c.}$ being the Hermitian
conjugate. Here, ${\hat\alpha}={\hat x},{\hat y}$ denotes the unit
vector in the $\alpha$-direction. The spinon ($b_{i\sigma}^\dagger,
b_{i\sigma},\sigma=\uparrow,\downarrow$) and holon
($h^\dagger_i,h_i$) operators are both bosonic.
The mutual statistics obeyed by spinons and holons is accounted for
the following topological constraints satisfied by the gauge fields.
For each loop ${\cal C}$ in the lattice plane,
\begin{eqnarray}
\begin{array}{c}
\sum_{x\in {\cal C}} A_\alpha^s J^\alpha_{\cal C} \equiv \pi
\sum_{x\,{\rm inside}\, {\cal C}} \left( b^\dagger_{\uparrow}
b_{\uparrow}-b^\dagger_{\downarrow} b_{\downarrow}\right)\,\, {\rm
mod}\,\, 2\pi, \\
\sum_{x\in {\cal C}} A_\alpha^h J^\alpha_{\cal C} \equiv \pi
\sum_{x\,{\rm inside}\, {\cal C}} h^\dagger h\,\, {\rm mod}\,\,
2\pi.
\end{array}
\label{plaque1}
\end{eqnarray}
Here, $J_{\cal C}^\alpha(x)= +1\, (-1)$ for the link $x \rightarrow
x+{\hat \alpha}\, (x+{\hat \alpha}\rightarrow x) $ on ${\cal C}$,
and is zero otherwise.

In the underdoped regime, it suffices to invoke the mean field
approximation \cite{Weng07,Weng97}, with the Hamiltonian simplified
to $H = -\sum_{i\alpha}(J_s \sum_\sigma e^{-i\sigma A_\alpha^h(i)}
b_{i-\sigma}^\dagger b_{i+\ha \sigma}^\dagger + t_h
e^{i(A_\alpha^s(i)+A_\alpha^e(i))} h_i^\dagger h_{i+\ha} ) + {\rm
h.c.}$ ($J_s, t_h>0$). The external gauge field $A^e_\alpha(i)$
generates an electromagnetic field and couples to the holon degree
of freedom. The Hamiltonian $H$ and the constraints (\ref{plaque1})
constitute our exact
starting point.

To proceed further, we consider the coherent state path integral
representation of the partition function, ${\cal Z}={\rm Tr}\,
e^{-\beta H}$, where $\beta\, (\rightarrow \infty)$ is the inverse
temperature. The technical challenge arises mainly from taking the
topological constraints (\ref{plaque1}) into account. Such a task
was fulfilled in Ref.~\cite{KQW05} in the continuum limit. Here, we
extend substantially the previous results to a more realistic
lattice model. We are able to show ${\cal Z}
= \int\!\!\!\!\int_D e^{-\sum_x
\mathscr{L}
}$ with
\begin{eqnarray}
\mathscr{L}&=& \mathcal{L}_{s}+\mathcal{L}_h+V \nonumber\\
&& + \frac{i}{\pi} \epsilon^{\mu\nu\lambda} \left(A_{{\mu}}^s - 2\pi
\mathscr{N}_\mu^{s}\right)d_\nu \left(A_{{\lambda}}^h - 2\pi
\mathscr{N}_\lambda^{h}\right)\label{compactaction}.
\end{eqnarray}
Here, $\mathscr{N}_\mu^{s,h}$ are integer fields and ${\cal
L}_{s,h}$ is the spinon and holon Lagrangian, respectively,
\begin{align*}
& {\cal L}_s = (b^\dagger_{\uparrow},b_{\downarrow}) \left(
                            \begin{array}{cc}
                              D^h_{0\uparrow} +\lambda^s & -J_s D_\alpha^h \\
                              -J_s D_\alpha^h & D_{0\downarrow}^h +\lambda^s \\
                            \end{array}
                          \right)
\left(
  \begin{array}{c}
    b_{\uparrow} \\
    b_{\downarrow}^\dagger \\
  \end{array}
\right), \cr & \qquad {\cal L}_h = h^\dagger\left(D_0^s + \lambda^h
- t_h D^s_\alpha \right)h.
\end{align*}
Here and below, the Einstein's summation convention is implied and
the summation over the indices $\mu,\nu,\lambda$ includes both the
imaginary time and spatial components. The on-site repulsive
potential $V$ softens the hard-core boson condition and depends on
$h^\dagger h$ and $b^\dagger_\sigma b_\sigma$. We shall not further
present its details to which the results below are insensitive.
$\epsilon^{\mu\nu\lambda}$ is the totally antisymmetric tensor.
Finally, the notation ``$\int\!\!\!\!\int_D$'' stands for the
integral over the fields: $b^\dagger,b,h^\dagger,h,A^{s,h}$ and the
Lagrange multipliers $\lambda^{s,h}$, and the summation over
$\mathscr{N}_\mu^{s,h}$, $D_\alpha^{s}=e^{i
(A_{\alpha}^{s}+A_{\alpha}^e)}e^{-d_\alpha}+e^{d_\alpha}e^{-i
(A_{\alpha}^{s}+A_{\alpha}^e)}$, $D_\alpha^{h}=e^{i
A_{\alpha}^{h}}e^{-d_\alpha}+e^{d_\alpha}e^{-i A_{\alpha}^{h}}$,
$D_0^{s}=d_0 - iA_0^{s}$, $D_{0\sigma}^{h}=\sigma (d_0 - i A_0^{h})$
with $d_\mu$ the lattice derivative. With these preparations, the
pair of order parameters are defined as
\begin{eqnarray}
W_\delta^{s,h}[{\cal C}]\equiv {\cal Z}^{-1} \int\!\!\!\!\int_D
e^{-\sum_x ( \mathscr{L} -i A^{s,h}_\mu J_{\cal C}^{\mu})},
\label{Wilson}
\end{eqnarray}
where ${\cal C}$ is a spacetime rectangle with length $T$ ($R$) in
the imaginary time (spatial) direction and $T\gg R$.

Differing from the prototypical field theory \cite{KQW05},
$\mathscr{L}$
keeps firm track of the compact nature of $A_\alpha^{s,h}$, which
affects profoundly the ground state properties, as will be shown
below. In particular, the {\it lattice mutual Chern-Simons term},
namely the last term in Eq.~(\ref{compactaction}) (Such a term was
found previously in a study of Josephson junction arrays
\cite{DST96}.), is periodic under a shift: $A^{s,h}_\alpha
\rightarrow A^{s,h}_\alpha + 2\pi m^{s,h}_\alpha,
\mathscr{N}^{s,h}_\alpha \rightarrow \mathscr{N}^{s,h}_\alpha + 2\pi
m^{s,h}_\alpha$ with $m^{s,h}_\alpha\in \Bbb{Z}$. Moreover, summing
up $\mathscr{N}_0^{s,h}$ enforces $\epsilon^{0\mu\nu}d_\mu
A_\nu^{h,s}$ to be $m\pi$ with $m\in \Bbb{Z}$.


{\it Superconducting phase.}--Consider the case of dilute spin
excitations where we may ignore the spinon field, i.e., $b^\dagger
=b =0$. Then, $\lambda^h |h|^2 + V$ gives rise to the holon
superfluid. More precisely, factorizing the holon field as $h(x) =
|h| e^{i\theta(x)}$ and inserting it into ${\cal L}_h+V$, we find
that the fluctuation of $|h|$ is massive, whereas the Goldstone mode
$\theta(x)$ is massless. Therefore, we ignore the terms associated
with the spatial fluctuations of $|h|$ and obtain ${\cal L}_h = i
|h|^2 (d_0 \theta-A_0^s -i\lambda^h) - 2 t_h |h|^2 \sum_\alpha \cos
(d_\alpha \theta - A_{\alpha}^s )$ in the absence of $A_{\alpha}^e$,
which is further simplified to $i |h|^2 (d_0 \theta-A_0^s
-i\lambda^h)
 + t_h |h|^2 [(d_\alpha
\theta - A_{\alpha}^s)^2-2]$. (By the definition of $H$, a
$2\pi$-shift in $d_\alpha\theta$ is absorbed into $A_\alpha^s$.)

To calculate $W_\delta^{h}[{\cal C}]$, we separate $A^h$ into the
background value and the fluctuation. The former leads to a uniform
flux, $\pi\delta$, at each plaquette and does not contribute to
$W_\delta^{h}[{\cal C}]$. Then, we introduce the unitary gauge so as
to incorporate $d_\mu\theta$ into $A_\mu^s$, and insert the
simplified expression of ${\cal L}_h$ into $\mathscr{L}$.
Integrating out the matter and $A^s$ fields, we find
\begin{eqnarray}
&& W_\delta^{h}[{\cal C}] \sim \sum_{\{\mathscr{N}^s\}} \int D(a^h)
e^{-\frac{1}{4} \sum_x F^{h}_{\mu\nu}F^{h\,\mu\nu}} \nonumber\\
&& \qquad \qquad \qquad \times e^{i \pi \sqrt{2 t_h \delta} \sum_x
a^h_\mu (J^\mu_{\cal C}+2 \epsilon^{\mu\nu\lambda} d_\nu
\mathscr{N}^s_\lambda)}
\label{eloop}
\end{eqnarray}
upon appropriate rescaling,
where $F^h_{\mu\nu}=d_\mu a^h_\nu-d_\nu a^h_\mu$ is the Maxwell
tensor with $a^h$ the fluctuating component of $A^h$, and $ \sqrt{2
t_h \delta}$ is the bare ``charge''. In the subsequent step, we
integrate out $a^h$ by using the Feynman gauge, which leads to
important consequences. First of all, we find that at the ground
state (where the external source $\pi J^\mu_{\cal C}$ is absent),
two phase vortices of the holon superfluid carrying opposite
vorticity $\pm 2\pi\epsilon^{0\mu\nu} d_\mu \mathscr{N}^s_\nu$ are
logarithmically confined. Therefore, no free phase vortices exist
with $\mathscr{N}^s_\alpha$ set to zero. Then, $\pi J^\mu_{\cal C}$
mimics an external dipole which may be produced by a pair of excited
spinons with identical or opposite spin polarizations. In the latter
case, a vortex with a vorticity of $-2\pi$ is excited from the
background and bound to a spinon, forming a dipole. Taking these
considerations into account, we find
\begin{eqnarray}
\ln W_\delta^{h}[{\cal C}] \sim - \pi t_h\delta\, T\ln
R,\label{eforce}
\end{eqnarray}
which shows that the spinons are logarithmically confined. To
calculate $W_\delta^{s}[{\cal C}]$, we ignore
$a_\alpha^h$.
Integrating out the matter fields, we find the holon deconfinement,
\begin{eqnarray}
\ln W_\delta^{s}[{\cal C}] &\sim& \ln \int D(A^s)\, e^{-t_h \delta
\sum_x A_\mu^s A^{s\mu}+i\sum_x
A^s_\mu J^\mu_{\cal C}} \nonumber\\
&\sim& -\frac{1}{4 t_h \delta} (T+R).
\label{mloop}
\end{eqnarray}

To further probe the SC long range order, we consider the response
to a static small magnetic field. For this purpose we need to
substitute $A^s_\alpha $ by $ A^s_\alpha +A^e_\alpha$ in ${\cal
L}_h$. If $A^e_\alpha=2m\pi$ ($m\in \Bbb{Z}$), the partition
function ${\cal Z}$ remains unchanged. For $A^e_\alpha=(2m+1)\pi$,
the additional $\pi$-phase may be absorbed into $A^s_\alpha$,
leaving ${\cal L}_h$ (and thereby ${\cal Z}$) unaffected. According
to Eq.~(\ref{plaque1}), we conclude that a single spin, with two
possible polarization directions, is locally excited and nucleated
at the magnetic vortex core \cite{Weng02}. As such, the external
magnetic field $m\pi$ flux is fully screened--a profound result of
the integer field $\mathscr{N}_0^h$. If the flux value is not
$m\pi$, the magnetic field is excluded by the superconductor.
Indeed, in this case,
${\cal Z}$ is merely determined by  $A^e $ modulo $\pi$ (denoted as
${\tilde A^e}$), ${\cal Z} \sim \int D(\theta)\, e^{2t_h \delta\beta
\sum_{i\alpha} \cos \left(d_\alpha \theta - {\tilde A^e}_\alpha
\right)} \sim e^{-t_h \delta\beta \sum_{{\bf q}}|{\tilde A^e}_\perp
({\bf q})|^2}$,
which justifies the Meissner effect. Here ${\tilde A^e}_\perp ({\bf
q})$ is the transverse part of the Fourier transform of ${\tilde
A^e}$.

{\it Antiferromagnetic phase.}--We turn now to the case where holons
are dilute and, likewise, set $h^\dagger =h =0$.
Then, $\lambda^s (|b_\uparrow|^2+|b_\downarrow|^2) + V$ gives rise
to a two-component spinon superfluid, with $|b_\uparrow|\approx
|b_\downarrow|\approx \sqrt {n}$ implying a magnetization in the
transverse direction. Here $n$ is the concentration of condensed
spinons depending on $\delta$. Similar to the discussions on the SC
phase, we factorize the two-component spinon field,
$(b^\dagger_\uparrow (x), b_\downarrow (x))$, as $(|b_\uparrow|,
|b_\downarrow|) e^{-i\theta(x)}$ and insert it into ${\cal L}_s+V$.
Ignoring the spatial fluctuations of $|b|$, we obtain ${\cal L}_s =
\sum_\sigma |b_\sigma|^2 [i(\sigma d_0 \theta- \sigma A_0^h -
i\lambda^s)- 2 J_s \sum_\alpha \cos (d_\alpha \theta - A_{\alpha}^h
)]$. Integrating out the matter and $A^h$ fields gives
\begin{eqnarray}
&& W_\delta^s[{\cal C}] \sim \sum_{\{\mathscr{N}^h\}} \int\!\!
D(A^s)
e^{- \frac{1}{4}\sum_x F^s_{\mu\nu}F^{s\,\mu\nu}} \nonumber\\
&& \qquad \qquad \times e^{i \pi \sqrt{4 n J_s} \sum_x A^s_\mu
(J^\mu_{\cal C}+2 \epsilon^{\mu\nu\lambda} d_\nu
\mathscr{N}^h_\lambda)},
\label{mloop1}
\end{eqnarray}
where $F^s_{\mu\nu}=d_\mu A^s_\nu-d_\nu A^s_\mu$.

Eq.~(\ref{mloop1}) has far-reaching consequences. First of all, by
integrating out the $A^s$ field, we find that at the ground state
($\pi J^\mu_{\cal C}=0$) the phase vortices of the spinon superfluid
carrying opposite vorticity $\pm 2\pi\epsilon^{0\mu\nu} d_\mu
\mathscr{N}^h_\nu$ are logarithmically confined. Most importantly,
$\pi J^\mu_{\cal C}$ mimics an external dipole produced by a pair of
the holon and anti-holon. The latter is a $-\pi$-fluxoid, formed out
of a $-2\pi$ phase vortex and a $\pi$-flux carried by the holon.
Such a holon-anti-holon pair is logarithmically confined,
\begin{eqnarray}
\ln W_\delta^{s}[{\cal C}] \sim - 2\pi J_s n\, T\ln R.
\label{mforce1}
\end{eqnarray}
In other words, a holon pair nucleates in a phase vortex of the
spinon superfluid of vorticity $-2\pi$, forming a ``neutral''
object. Furthermore, since the phase vortex is static, the pair of
holons is spatially localized and the AF phase is insulating (see
below for further explanations). It should be noticed that without
the integer field $\mathscr{N}^h_\mu$, such an insulating phase
cannot be established. Instead, the SC phase is pushed all the way
to $\delta=0$ \cite{Weng07}. Repeating the derivation of
Eq.~(\ref{mloop}), we find the spinon deconfinement,
\begin{eqnarray}
\ln W_\delta^{h}[{\cal C}] \sim -\frac{1}{8 J_s n} (T+R).
\label{eloop1}
\end{eqnarray}

{\it Bose insulating phase.}--The analytic results, namely
Eqs.~(\ref{eforce}), (\ref{mloop}), (\ref{mforce1}) and
(\ref{eloop1}), allow us to make an important observation. The
asymptotic behavior, Eqs.~(\ref{eforce}) and (\ref{eloop1}), signal
a critical concentration separating the spinon confinement and
deconfinement phases, at which $W_\delta^{h}[{\cal C}]$ is
nonanalytic in $\delta$. Indeed, in the SC phase, the spinon
excitations become progressively important as $\delta$ decreases:
they cause a renormalization of the bare ``charge''
in Eq.~(\ref{eforce}) and eventually drive the system to the spinon
deconfinement QCP where the ``charge'' vanishes. For smaller
$\delta$ the SC long range order disappears.
Likewise, Eqs.~(\ref{mloop}) and (\ref{mforce1}) signal another
critical concentration, at which $W_\delta^{s}[{\cal C}]$ is
nonanalytic, separating the holon confinement and deconfinement
phases. Renormalizing the bare ``charge'' in Eq.~(\ref{mforce1})
drives the system towards this QCP. For $\delta$ larger than this
critical value, the AF long range order disappears.
Because these two renormalization mechanisms are independent, these
two QCPs are generally not identical, giving an intermediate phase
where both the AF and SC long range orders vanish and both the holon
and spinon are deconfined (condensed).

This intermediate phase does not support dc electric transport. To
prove this, we notice that the composition rule (\ref{conductivity})
is valid for the entire underdoped regime. Indeed, minimizing
$\mathscr{L}$ gives $\frac{\delta {\cal L}_{s,h}}{\delta
A^{h,s}_\alpha}=-\frac{i}{\pi} \epsilon^{\alpha\mu\nu} d_\mu
(A^{s,h}_\nu-2\pi \mathscr{N}^{s,h}_\nu)$, i.e.,
$j^{s,h}_\alpha=\frac{1}{\pi} \epsilon^{0\alpha\beta}
E^{s,h}_\beta$. By definition, $j_\alpha^s\equiv \sigma_sE^h_\alpha$
and $j_\alpha^h\equiv\sigma_h (E_\alpha^s+E_\alpha)$. Noticing that
the electric and holon currents are identical, we obtain
Eq.~(\ref{conductivity}) from these three relations. The striking
structure of the composition rule is a consequence of i) that the
two pieces involved are vortices, and ii) that they obey mutual
statistical interaction. (This fact has been established in a
completely different context \cite{Galitskii05}.)
For the BI phase, $\sigma_s\rightarrow \infty$ because of spinon
condensation, implying $\sigma=0$.

{\it Crossover from Mott's law to activation law.}--Finally, we
discuss a possible experimental observation of resistivity in the AF
phase at sufficiently low but nonvanishing temperatures. In this
regime, a holon-anti-holon pair bound to the vortex of spinon
superfluids displays two-dimensional variable-range hopping
conduction known as the Mott's law, $\ln \sigma_h^{-1}\sim T^{-1/3}$
\cite{Shklovskii}. On the other hand, spinon superfluid is actually
the so-called strongly entangled vortex phase \cite{Larkin93}.
Carrying the arguments of Ref.~\cite{Larkin93} to the present
context, we find that the spin transport, in response to ${\bf
E}^h$, displays an activation law, i.e., $\ln\sigma_s\sim T^{-1}$.
According to Eq.~(\ref{conductivity}), the resistivity displays a
crossover from Mott's law to activation law upon decreasing
temperatures, which may serve as a probe of holon (spinon)
confinement (deconfinement) in the AF phase.

We thank J. Zaanen for discussions and M. Garst, V. Gurarie, and
L.-H. Tang for conversations. Work supported by NSFC Nos. 10688401
and 10834003, by MOST National Program for Basic Research, by DFG
SFB/TR12, and by HK RGC No. HKUST3/CRF/09.


\begin{thebibliography}{}

\bibitem{Anderson87}
Z. Zou and P. W. Anderson, Phys. Rev. B
\textbf{37}, 627 (1988).


\bibitem{Lee06} For a review, see, P. A. Lee, N. Nagaosa, and X. G. Wen, Rev. Mod. Phys. \textbf{78},
17 (2006).

\bibitem{Fisher04} T. Senthil, {\it et. al.},
Science \textbf{303}, 1490 (2004).



\bibitem{Fisher00} T. Senthil and M. P. A. Fisher,
Phys. Rev. B \textbf{62}, 7850 (2000).

\bibitem{Sachdev03} S. Sachdev, Rev. Mod. Phys. \textbf{75},
913 (2003).

\bibitem{Zaanen09} J. Zaanen and B. J. Overbosch, arXiv:0911.4070.

\bibitem{Weng97} Z. Y. Weng, {\it et. al.},
Phys. Rev. B \textbf{55}, 3894 (1997).

\bibitem{Weng07} See, for a review, Z. Y. Weng, Intl. J. Mod. Phys. B \textbf{21},
773 (2007).

\bibitem{KQW05} S. P. Kou, X. L. Qi, and Z. Y. Weng, Phys. Rev. B
\textbf{71}, 235102 (2005).

\bibitem{Weng02} V. N. Muthukumar and Z. Y. Weng, Phys. Rev. B \textbf{65}, 174511 (2002).

\bibitem{Weng10} J. W. Mei and Z. Y. Weng, Phys. Rev. B \textbf{81}, 014507 (2010).

\bibitem{Xu09} C. Xu, Phys. Rev. B \textbf{83}, 024408 (2011).

\bibitem{unit} The spacetime lattice constant, the Planck's constant, the speed of light
in vacuum, and the electron charge are set to unity.

\bibitem{Galitskii05} V. M. Galitskii, {\it et. al.}
Phys. Rev. Lett. \textbf{95}, 077002 (2005).





\bibitem{DST96} M. C. Diamantini, {\it et. al.},
Nucl. Phys. B \textbf{474}, 641 (1996).

\bibitem{Shklovskii} B. I. Shklovskii and A. L. Efros, {\it Electronic properties of doped semiconductors}
(Springer, Berlin, 1984).

\bibitem{Larkin93} M. V. Feigelman, {\it et. al.}
Phys. Rev. B \textbf{48}, 16641 (1993).



\end{thebibliography}
\end{document}